\documentclass[10pt,conference]{IEEEtran}
\usepackage{amsmath, amssymb, bm, cite, epsfig, psfrag}
\usepackage{graphicx}
\usepackage[margin=0.7in]{geometry}
\usepackage{dblfloatfix}
\usepackage{array}
\usepackage{textcomp}
\newcolumntype{P}[1]{>{\centering\hspace{0pt}}p{#1}}
\newcolumntype{M}[1]{>{\centering\hspace{0pt}}m{#1}}
\newcolumntype{L}{>{\centering\arraybackslash}m{3cm}}
\usepackage{longtable}
\usepackage{supertabular,booktabs}
\usepackage{multirow}
\usepackage[usenames,dvipsnames]{xcolor}

\usepackage{etoolbox}
\usepackage{pbox}
\usepackage{fixltx2e}
\usepackage{romannum}
\usepackage{filecontents}
\usepackage{enumerate}
\graphicspath{ {figures/} }
\usepackage{subfigure}
\usepackage{colortbl}
\usepackage{fancyhdr}

\usepackage{bm}
\newtoggle{conference}
\togglefalse{conference} %
\graphicspath{{figures/}}

\fancypagestyle{firststyle}{
	\fancyhf{}
	\fancyhead[L]{ O. Kanhere, S. Ju, Y. Xing, and T. S. Rappaport, ``Map-Assisted Millimeter Wave Localization for Accurate Position Location,'' \textit{in GLOBECOM 2019 - 2019 IEEE Global Communications Conference,} Hawaii, U.S.A, Dec. 2019, pp. 1--6.}     %

}
\usepackage{etoolbox}
\makeatletter
\patchcmd{\@makecaption}
{\scshape}
{}
{}
{}
\makeatletter
\patchcmd{\@makecaption}
{\\}
{.\ }
{}
{}
\makeatother

\begin{document}
\title{Map-Assisted Millimeter Wave Localization for Accurate Position Location}
\author{\IEEEauthorblockN{Ojas Kanhere, Shihao Ju, Yunchou Xing, Theodore S. Rappaport\\}
	
\IEEEauthorblockA{	\small NYU WIRELESS\\
					NYU Tandon School of Engineering\\
					Brooklyn, NY 11201\\
					\{ojask,shao,ychou,tsr\}@nyu.edu}}\vspace{-0.7cm}

\maketitle
\thispagestyle{firststyle}

\begin{abstract}

Accurate precise positioning at millimeter wave frequencies is possible due to the large available bandwidth that permits precise on-the-fly time of flight measurements using conventional air interface standards. In addition, narrow antenna beamwidths may be used to determine the angles of arrival and departure of the multipath components between the base station and mobile users. By combining accurate temporal and angular information of multipath components with a 3-D map of the environment (that may be built by each user or downloaded a-priori), robust localization is possible, even in non-line-of-sight environments. In this work, we develop an accurate 3-D ray tracer for an indoor office environment and demonstrate how the fusion of angle of departure and time of flight information in concert with a 3-D map of a typical large office environment provides a mean accuracy of 12.6 cm in line-of-sight and 16.3 cm in non-line-of-sight, over 100 receiver distances ranging from 1.5 m to 24.5 m using a single base station. We show how increasing the number of base stations improves the average non-line-of-sight position location accuracy to 5.5 cm at 21 locations with a maximum propagation distance of 24.5 m.

\end{abstract}
    

\section{Introduction}\label{Introduction}
Positioning is the determination of the location of a user that is fixed or moving, based on the known locations of other reference points, which are usually base stations (BSs). In line-of-sight (LOS) environments, the time of flight (ToF) of a signal can be used to estimate the transmitter (TX) - receiver (RX) separation distance $ d $ (since $ d = c\cdot t $, where $ c $ is the speed of light and $ t $ is the ToF). The RX's position may then be estimated by trilateration \cite{Kanhere18a}. In non-LOS (NLOS) environments, ranging based on ToF alone introduces a positive bias in the position estimates since the path length of reflected multipath rays is longer than the true distance between the user and the BS.

The ultra-wide bandwidths available at mmWave and Terahertz (THz) frequencies allow the RXs to resolve finely spaced multipath components and accurately measure the ToF of the signal \cite{Mac17a,Rappaport19a}. The phase accrued by a signal in LOS is proportionate to the ToF \cite{Kanhere18a}. In \cite{Parker16}, the phase of the received signal in LOS was used to estimate the TX-RX (TR) separation distance in LOS at 300 GHz. The phase of a signal rotates $ 2\pi $ radians every $ \lambda $ meters \cite{Rap02a}, yet phase ambiguity arises, since signals that traverse distances that differ by integral multiples of $ \lambda $ incur the same phase. By tracking the phase of the transmitted signal at the RX and manually correcting for phase ambiguity, decimeter-level accuracy at 300 GHz was achieved up to distances of 40 m in LOS\cite{Parker16}.

\begin{figure}
	\centering
	\includegraphics[width=0.4\textwidth]{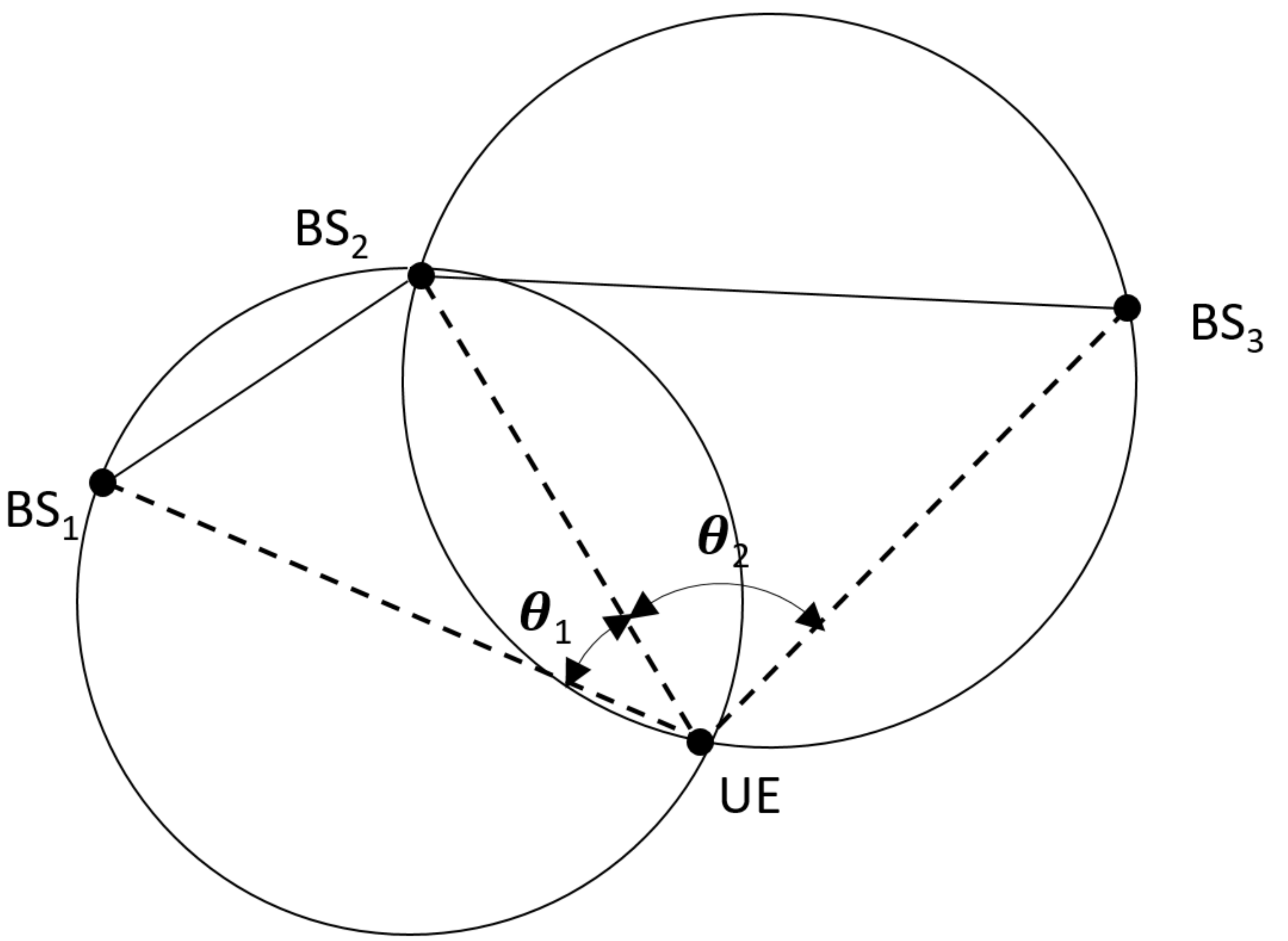}
	\caption{ The UE measures the relative AoAs, $ \theta_1 $ and $ \theta_2 $ from $ BS_1$, $BS_2$, and $BS_3 $. The UE is localized to the point where the circles drawn through the BSs (pairwise) and the UE intersect \cite{Rappaport88a,Rap89b,Bowditch80}.}
	\label{fig:angle_geometry}
		\vspace{-1.5em}
\end{figure}

The basic problem of determination of a mobile user's precise location from angular measurements is identical to the `three point problem' as known in land surveying \cite{Rappaport88a,Rap89b,Bowditch80} and may be used by electrically steerable phased array antennas in wireless systems at mmWave and THz frequencies. As seen in Fig. \ref{fig:angle_geometry}, relative angles between BSs are measured by the user. The unknown position of the mobile user is then calculated via basic trigonometry \cite{Rappaport88a,Rap89b}. Alternatively, as seen in Fig. \ref{fig:angle_geometry}, simple geometry may be used to observe that the locus of points where $ \text{BS}_1 $ and $\text{BS}_2 $ subtend a fixed angle $ \theta_1 $ is a circle circumscribing the triangle formed by the BSs (taken pairwise) and the user. The user location corresponds to the intersection of the two circles corresponding to the two relative AoAs measured between two BS pairs \cite{Rappaport88a,Rap89b}. The solution to the three point problem is sensitive to small errors in measured angles when either the BSs subtend a small angle at the user location, or when the observation point is on or near a circle which contains the three BSs. Look-up tables may also be used to localize the user, where the table stores the relative AoAs measured by the user at each location in the surveyed environment.

Narrow antenna beamwidths of antenna arrays at mmWave frequencies will allow the user device to accurately determine the AoA. In order to improve the AoA accuracy, two photodetectors were placed at an offset and mechanically rotated in \cite{Rappaport88a,Rap89b}. The sum and difference of the signals received by the two photodetectors was calculated to determine the precise angle, even with a course beam pattern. When the sum of the signals received at the two detectors exceeded a threshold and the absolute value of the difference in the signals fell below a threshold, the beacon was detected to be aligned with the photodetectors. An on-the-fly angular resolution of 0.2\textdegree~was achieved in \cite{Rappaport88a,Rap89b} with wideband photodetectors. The sum-and-difference method may also be implemented in an identical fashion by electrically steering an antenna array, using adjacent antenna beams or slightly offset antenna arrays with overlapping antenna patterns. A single antenna may also be used to locate the AoA of the peak signal from the BSs by quickly dithering the antenna boresight (electrically or mechanically). The sum and difference of the received signals at successive time instants may be used in place of measurements from two offset antennas.

In NLOS environments, due to specular reflections from walls and metallic surfaces, rays do not arrive from the direction of the BS, leading to accuracy penalties \cite{Kanhere18a} if used the AoA is used directly. Real-time electric beam steering algorithms facilitate scanning of room features in a matter of seconds. As a result, mobile phones of the future will likely be able to generate detailed 3-D maps on the fly, or will download them from the cloud \cite{Rappaport19a}. NLOS objects (around corners) may be ``viewed" by first rapidly scanning the environment via beam steering, in order to determine all the reflecting obstructions in the surroundings. The reflecting obstructions can then be distinguished from the target NLOS object to be ``viewed" by taking advantage of the fine temporal resolution at mmWave and sub-THz frequencies to create a 3-D map of the local environment \cite{Rappaport19a, Chi17a, Doddalla2018}. The 3-D maps may be utilized (in conjugation with angle of departure (AoD) from the known BSs and ToF measurements) to calculate, back-solve or estimate the actual paths that the multipath components take to reach the user as shown in Section \ref{sec:candidate_locations}. The paths taken by the multipath components that reach the user contain sufficient information to localize a user in NLOS, even in the absence of LOS multipath signals.

The remainder of this paper is organized as follows. Section \ref{sec:ray_tracer} describes NYURay, a 3-D mmWave ray tracer calibrated based on real-world measurements \cite{Mac15b}. Section \ref{sec:map_based} describes prior map-based localization algorithms described in literature. A new map-based localization algorithm, capable of centimeter-level localization using a single BS in LOS and NLOS environments is presented in Section \ref{sec:candidate_locations}. Simulations and results of the localization algorithm are presented in \ref{sec:simulations}, along with an analysis of position location accuracy based on the effect of varying the number of BSs used. The simulations were conducted using NYURay. Concluding remarks and directions for future work are provided in Section \ref{sec:conclusion}.
\section{NYURay - a 3-D mmWave Ray Tracer}\label{sec:ray_tracer}

Since wideband directional measurements are expensive and time consuming, a ray-tracer that is truthful to actual measurements at a wide range of locations is a powerful tool for exploring position location algorithms, data fusion, and overall position location accuracy and sensitivity. 

Building on the 2-D mmWave ray tracer developed in \cite{Kanhere18a}, NYURay, a 3-D mmWave ray tracer has been developed. NYURay is a hybrid ray tracer which combines shooting-bouncing rays (SBR) \cite{Schaubach92a, Durgin97a} with geometry-based ray tracing \cite{McKown91a,Ho94a}. 

A SBR ray tracer launches rays uniformly in all directions and then traces the path of each launched ray, as the ray interacts with various obstructions in the environment. Each launched ray represents a continuous wavefront. Each ray carries the power that would have been carried by the wavefront \cite{Durgin97a}. 

The accuracy of the AoA of rays received at the RX depends on the number of rays launched from the TX. For example, if rays are launched from the vertices of a tessellated icosahedron with tessellation factor $ N $, since the average radial separation between two rays is $ \dfrac{69^\circ}{N} $ \cite{Durgin97a}, for sub-degree accuracy for AoA, $ N>50 $, which is computationally expensive.

Image-based ray tracing \cite{McKown91a,Ho94a} relies on the principle that the incident ray and the reflected ray make the same angle with the normal to the plane containing the obstruction, which is computationally much easier. Obstructions are treated as infinitely long, thin mirrors when using image-based ray tracing \cite{Ho94a}. The image of the RX is taken, successively, in at most $ k $ obstructions, where $ k $ is the maximum number of reflections a ray may go through\cite{Ho94a}.

The shortcoming of image-based ray tracing is that if there are a large number of obstructions, the simulation run-time is large. Assuming that each ray is reflected at most three times, with $ N $ obstructions in the environment, there are ${N\choose 3}$ images that need to be computed ($ \mathcal{O}(N^3) $). Although the image-based ray tracing method finds the direction of arrival of rays very accurately, finding the reflection of the RX, recursively, from all combinations of obstructions is computationally expensive. 

\begin{figure}[]
	\centering
	\includegraphics[width=0.4\textwidth]{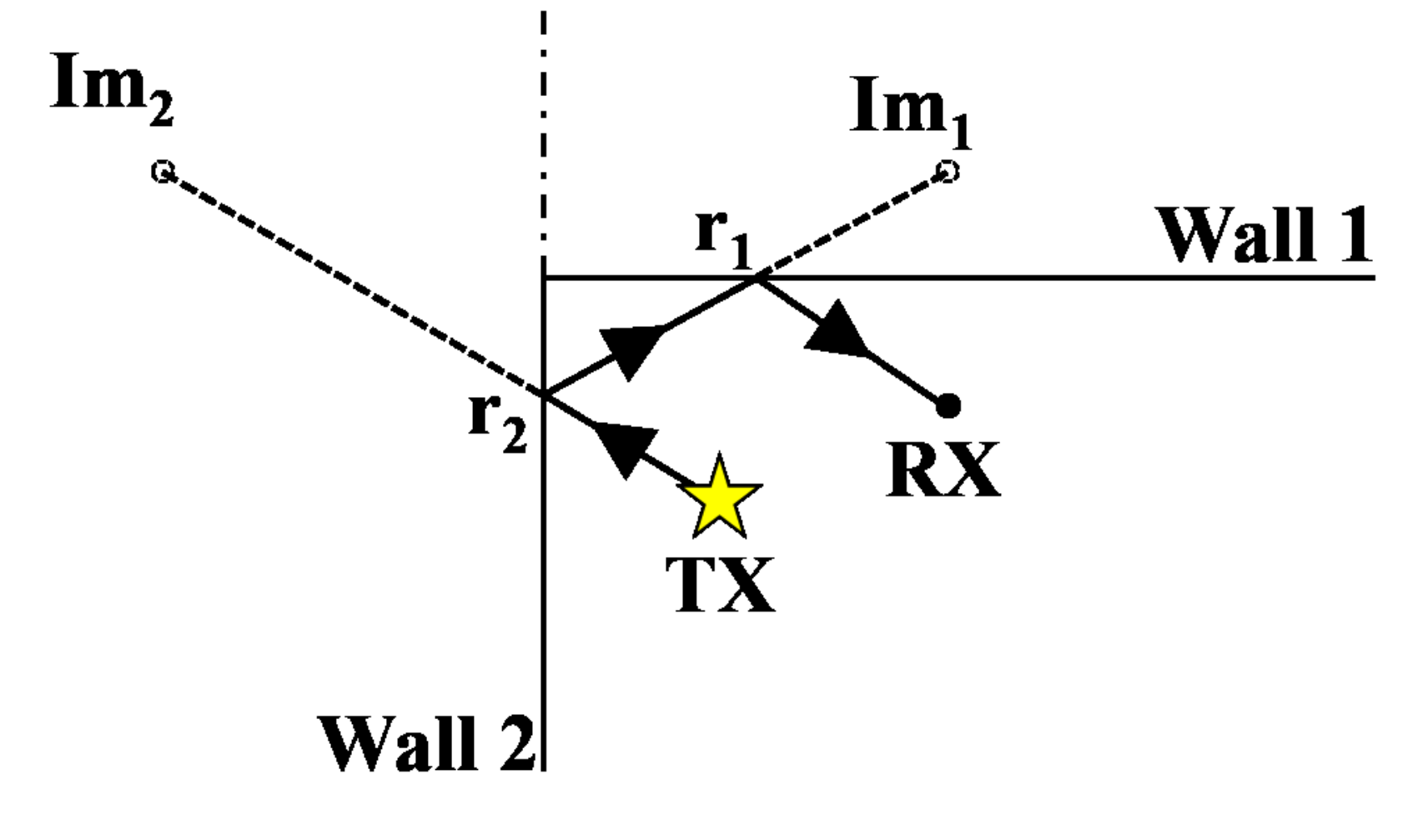}
	\caption{ In ray tracing based on images, successive images of the RX are found. Consider the successive images $ Im_1 $ and $ Im_2 $ of RX. $ Im_1 $ is the image of RX in Wall 1 and $ Im_2 $ is the images of $ Im_1 $ in Wall 2.}
	\label{fig:image_based}
\end{figure}

To reduce computational overhead, NYURay uses a hybrid ray tracing algorithm \cite{Tan96a}. The approximate trajectories of rays that reach the RX are first determined via SBR ray tracing. Once all the reflecting surfaces in the path of a ray are determined, image-based ray tracing is used to calculate the recursive reflections of the RX. The ray trajectory is accurately calculated by connecting all the RX images.

In every direction where a ray was launched, on encountering an intersection with an obstruction, two new rays were created - the specular reflected ray and the transmitted ray. By Snell's Law, the reflected ray and the incident ray make equal angles with the normal to the obstruction. The transmitted ray was assumed to propagate in the same direction as the incident ray. A linear model was used to characterize the variation of reflection coefficient $ \Gamma $ with incident angle $ \theta_i $, based on reflection measurements conducted in \cite{Xing19a,Ju19a} as follows:
\begin{align}
|\Gamma| = \dfrac{E_r}{E_i} =0.56\cdot  \theta_i  + 0.096,
\end{align}
where $ E_r $ is the reflected electric field, $ E_i $ is the incident electric field, and $ \theta_i$ is the angle of incidence of the ray. As a result, the reflected power $ P_r = |\Gamma|^2 P_i$, where $ P_i $ is the power incident on the obstruction. A constant transmission loss of 7.2 dB was assumed, based on the propagation measurements conducted in \cite{Mac15b}.  %Fig. \ref{fig:material_coefficients} illustrates how the $ |\Gamma|^2$ of a material varies with incident angle $ \theta_i $. 

New source rays at each boundary were then recursively traced in the reflection and transmission directions to the next encountered obstruction on the propagating ray path. Path loss was calculated based on the free space path loss (FSPL) model \cite{Sun16c}, with a TR separation distance equal to the total propagated ray length. Additionally, at most three reflections of rays were considered in order to reduce computation time. The limitation on the number of reflections was further justified by the observation that typically mmWave signals typically do not experience more than two reflections \cite{Rap13a}.

\begin{figure}
	\centering
	\includegraphics[width=0.5\textwidth]{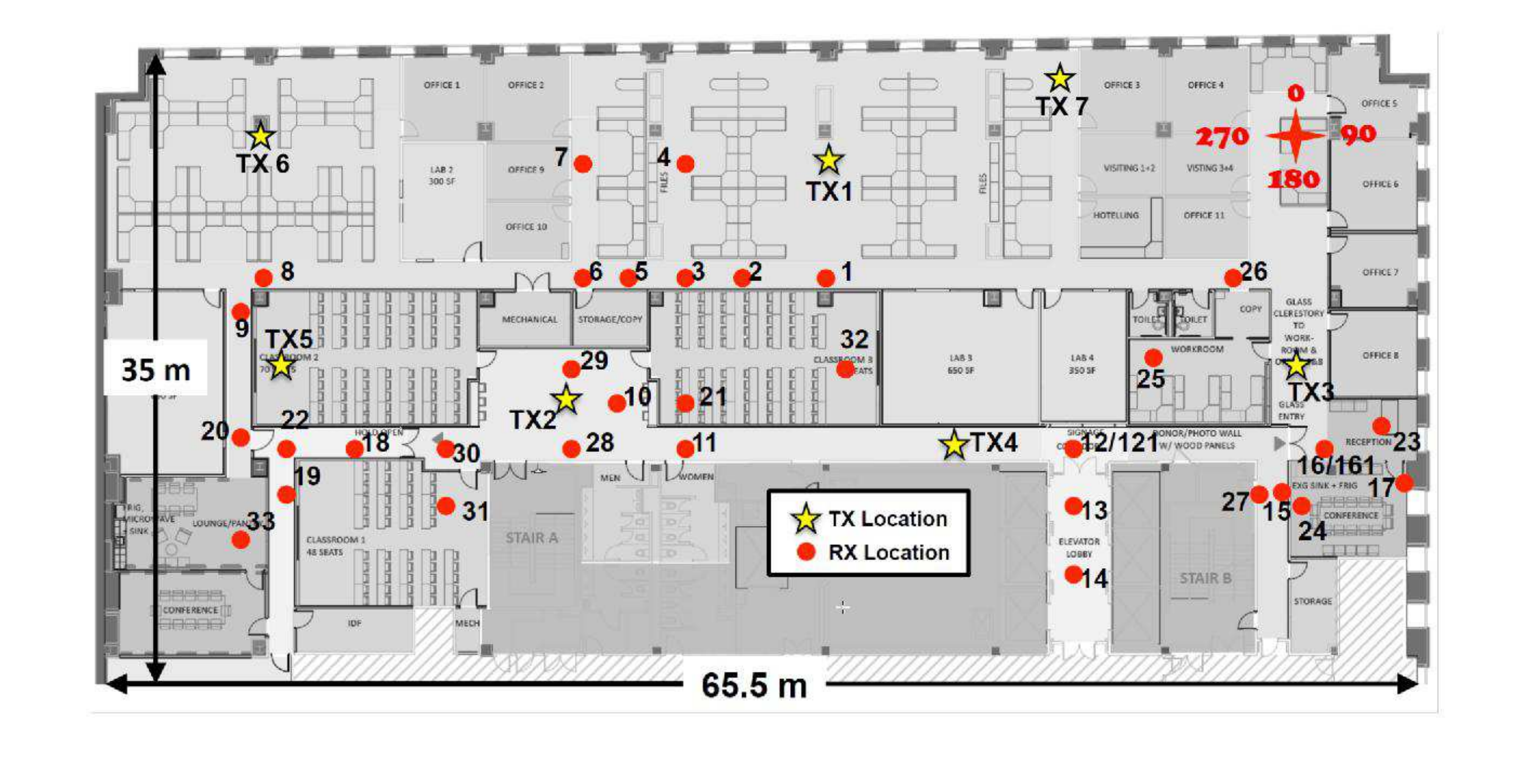}
	\caption{ Map of the 9th floor of 2 MTC, depicting the indoor locations where propagation measurements were conducted at 28 GHz and 73 GHz\cite{Mac15b}}
	\label{fig:Indoor_locations}
\end{figure}

Propagation measurements were conducted at 28 GHz and 73 GHz at the NYU WIRELESS research center, located on the 9th floor of 2 MTC using a 400 Megachip-per-second (Mcps) wideband sliding correlator channel sounder with high gain steerable antennas \cite{Mac15b}. The directional antennas used in the propagation measurements had antenna beamwidths of 30\textdegree and 15\textdegree, at 28 GHz and 73 GHz respectively. Fig. \ref{fig:Indoor_locations} depicts the five TX locations (TX 1-5) and the RX locations where the propagation measurements were conducted in \cite{Mac15b}. No measurements were conducted at TX locations 6 and 7, which were added during simulations to ensure adequate coverage throughout the floor.

\begin{figure}
	\centering
	\includegraphics[width=0.4\textwidth]{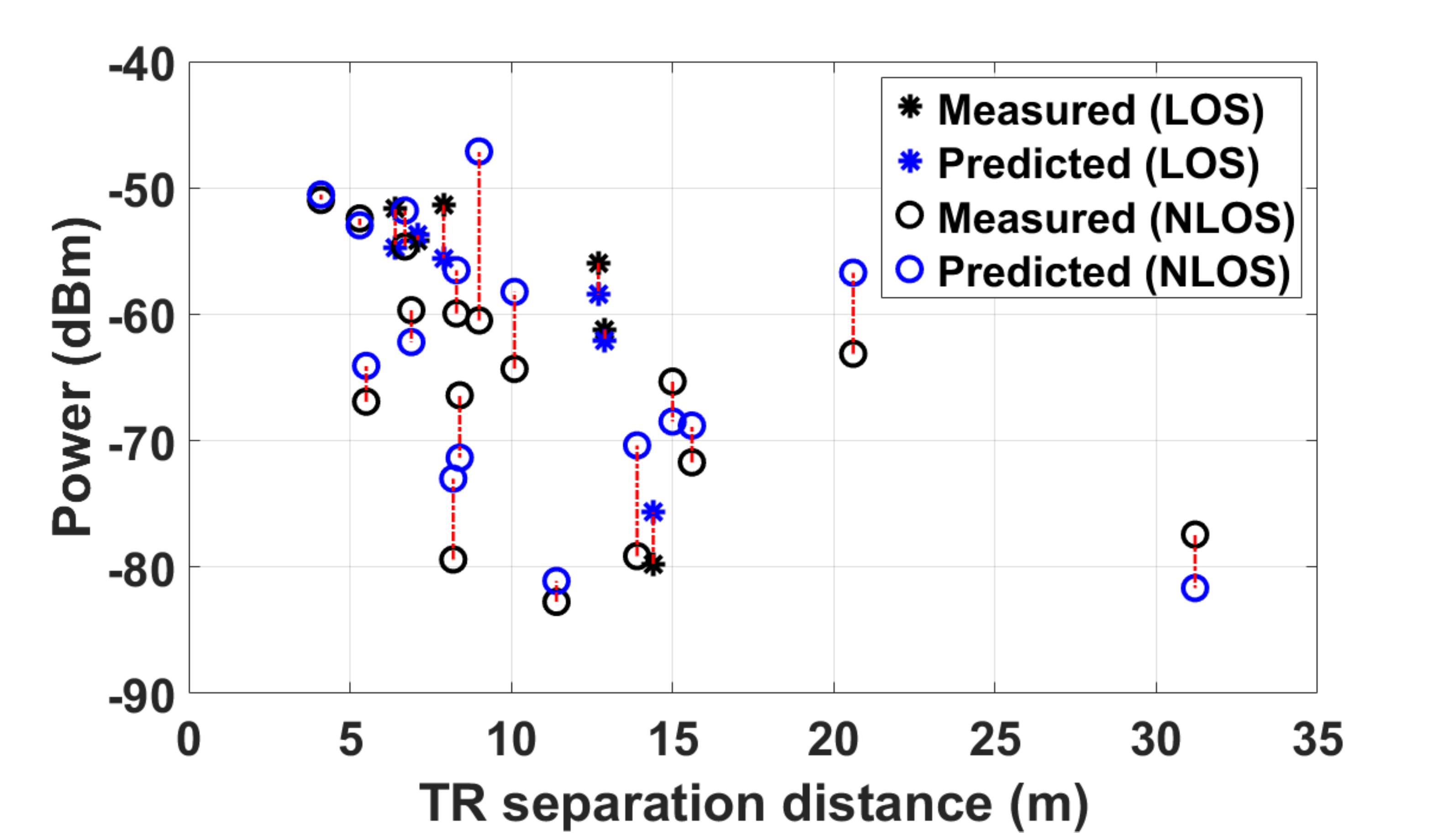}
	\caption{ Comparison between the measured and predicted powers using NYURay for 22 TX-RX links, distributed throughout the 9th floor of 2 MTC.}
	\label{fig:predicted_power}
\end{figure}

The performance of NYURay was evaluated by comparing the variation in the predicted and measured path loss with TR separation distance, as is illustrated in Fig. \ref{fig:predicted_power}. The 22 TX-RX links that were chosen for the comparison were spread throughout NYUWIRELESS. The mean of the received power prediction error was 0.41 dB and the standard deviation was 6.4 dB.

\section{Map-based localization }\label{sec:map_based}

In a multipath rich environment, source rays arrive at the RX via a direct path (if the direct path exists) as well as paths along which the source rays suffers multiple reflections. With knowledge of the angles at which the source rays arrive at the RX, the ToF of the source rays and a 3-D map of the surrounding environment, the RX may determine the location of the source. The rays along each AoA may be backpropagated, as though the RX were emitting the rays. Along the back-propagating path, the ray interacts with obstructions in a manner identical to how a forward-propagating ray would interact. 

In \cite{Kaya07a}, the intersection of two or more back-traced rays were labeled as candidate TX locations. The TX was localized to the weighted sum of all candidate TX locations. The candidates that had several other intersections in their vicinity were given greater weight. Using six RX locations, assuming perfect AoA resolution, a 90 percentile localization error of 5 m was achieved. The thermal noise floor was -85 dBm and a 30 dBm, 300 MHz signal was transmitted by the TX. 

Work in \cite{Aladsani19a} demonstrated how ToF, AoA measurements, along with a map of the obstacles encountered by the signal could be used to localize a user. A 3-D image of the environment was obtained via holographic 3-D imaging. A plane uniform linear array of antennas at the TX was used to calculate the signal scattered from objects in the environment, $ R(x,y,f) $, measured at point $ (x,y,z) $ using a vector network analyzer (VNA) at frequency $ f $, varying from 220 to 300 GHz. The image of the environment $ f(x,y,z) $ is given by: 
\begin{eqnarray}
f(x,y,z) =  IFT_{3-D} \left\{FT_{2D}\left\{R(x,y,f)\right\}\right\},
\end{eqnarray}
where IFT and FT are the discrete spatial inverse Fourier transform and Fourier transform respectively. Once a 3-D image of the environment was generated, the ToF and AoA of signals reaching the RX were estimated via beamforming techniques, as described in Section \ref{Introduction}. In \cite{Aladsani19a}, the authors assumed that all materials act like perfect reflectors and no rays penetrated the obstructions. Once the total back-traced ray length was equal to distance $ d $, the RX was localized to the end of the ray. Over a distance of 2.78 m, the proof of concept built in \cite{Aladsani19a} had an error of 2.6 cm. Although work in \cite{Aladsani19a} assumed that all obstructions were perfect reflectors at 300 GHz, measurements conducted in \cite{Rappaport19a} illustrate that at 140 GHz, the partition loss through glass and dry wall is close to 10 dB, indicating that materials do not act like perfect reflectors at mmWave and sub-THz frequencies.

The authors of \cite{Meissner13a} used ultra wideband (UWB) signals and estimates of ToF, with an indoor map for localization, using the principal of ``virtual access points" (VAs). Four TXs were used to localize the user in an area of approximately 20 m $ \times $ 5 m. The TX position was mirrored across each obstruction in the environment to create VAs. The VAs acted as additional anchor points to localize users. Centimeter level accuracy was achieved: using a signal with pulse width $ T_p = 0.2$ ns, a root mean squared (rms) positioning error of 3.2 cm was achieved. However \cite{Meissner13a} did not use AoA information, which will be available with high accuracy using directional antennas at mmWave frequencies.

\section{Localization with one or more BSs }\label{sec:candidate_locations}
Map-assisted positioning with angle and time (MAP-AT) is a novel map-assisted localization technique that shall now be presented. Localization is possible using a single BS, when at least two multipath components arrive at the user. The user need not be in LOS of the BS. MAP-AT can handle two types of BS-user configurations: the user may either be the TX or the RX of a known radio signal. If the user is the TX, ToF and AoA information is required. If the user is the RX, ToF and AoD information is required. Temporal and angular measurements impose constraints on the possible locations of a user. A 3-D map of the indoor environment imposes additional constraints on the user's location. These constraints shall be explained in the following sections.

\subsection{Configuration-\Romannum{1} - user in the reception mode}
In configuration-\Romannum{1}, the user receives a known signal from the BS. The BS may calculate the AoD of each multipath signal that reach the user during initial access. The ToF of each multipath component from the BS arriving at the user may either be estimated via the round trip time of the multipath component or the one-way propagation time. The BS may then send the AoD and ToF of each multipath component to the user via a feedback channel.

\begin{figure}
	\centering
	\includegraphics[width=0.35\textwidth]{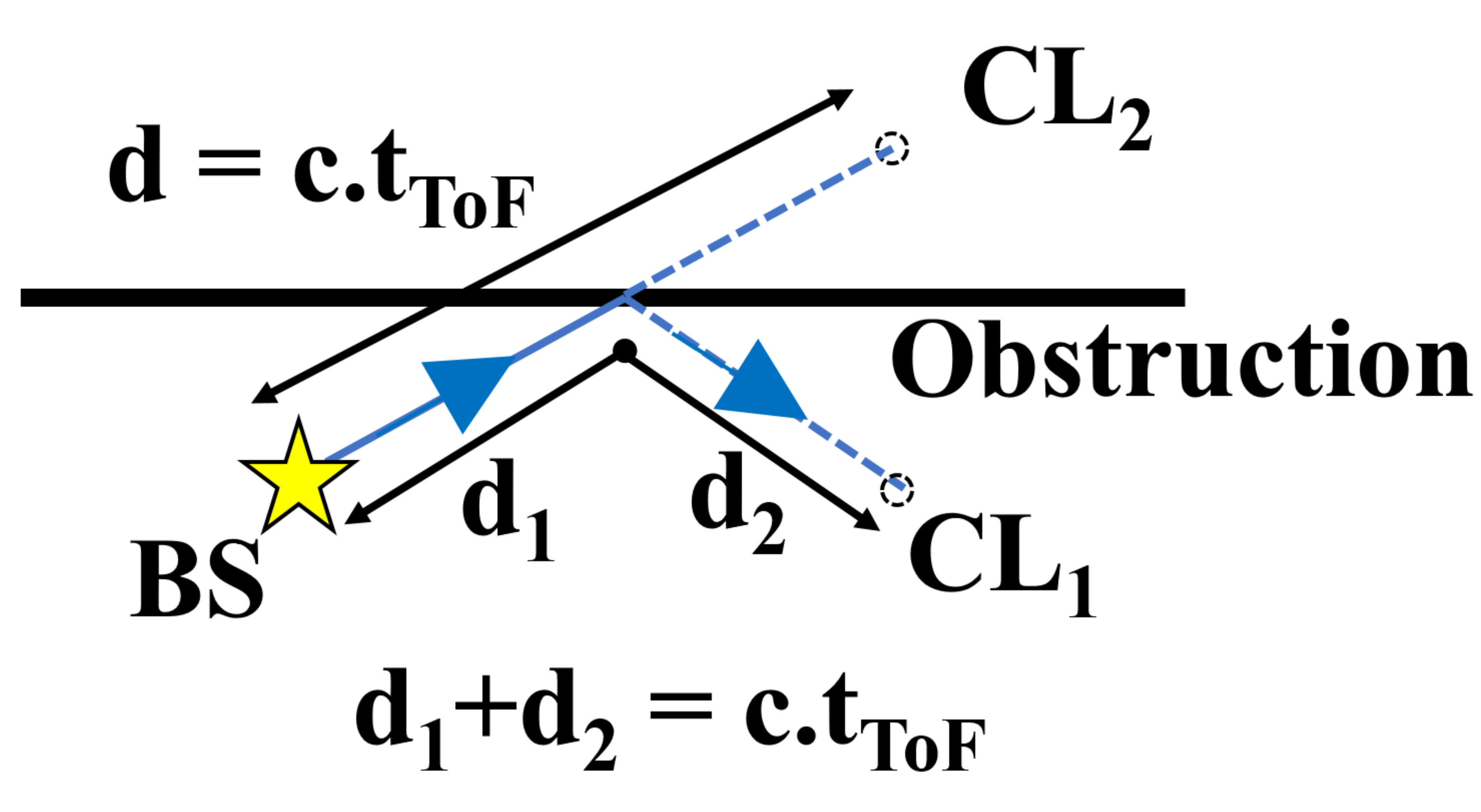}
	\caption{Configuration-\Romannum{1} : The two candidate locations of a user, $ \text{CL}_1$ and $\text{CL}_2 $, correspond to the cases where the signal sent from the BS is reflected by and passes through the obstruction, respectively.}
	\label{fig:config_1}
\end{figure}

\begin{figure}
	\centering
	\includegraphics[width=0.35\textwidth]{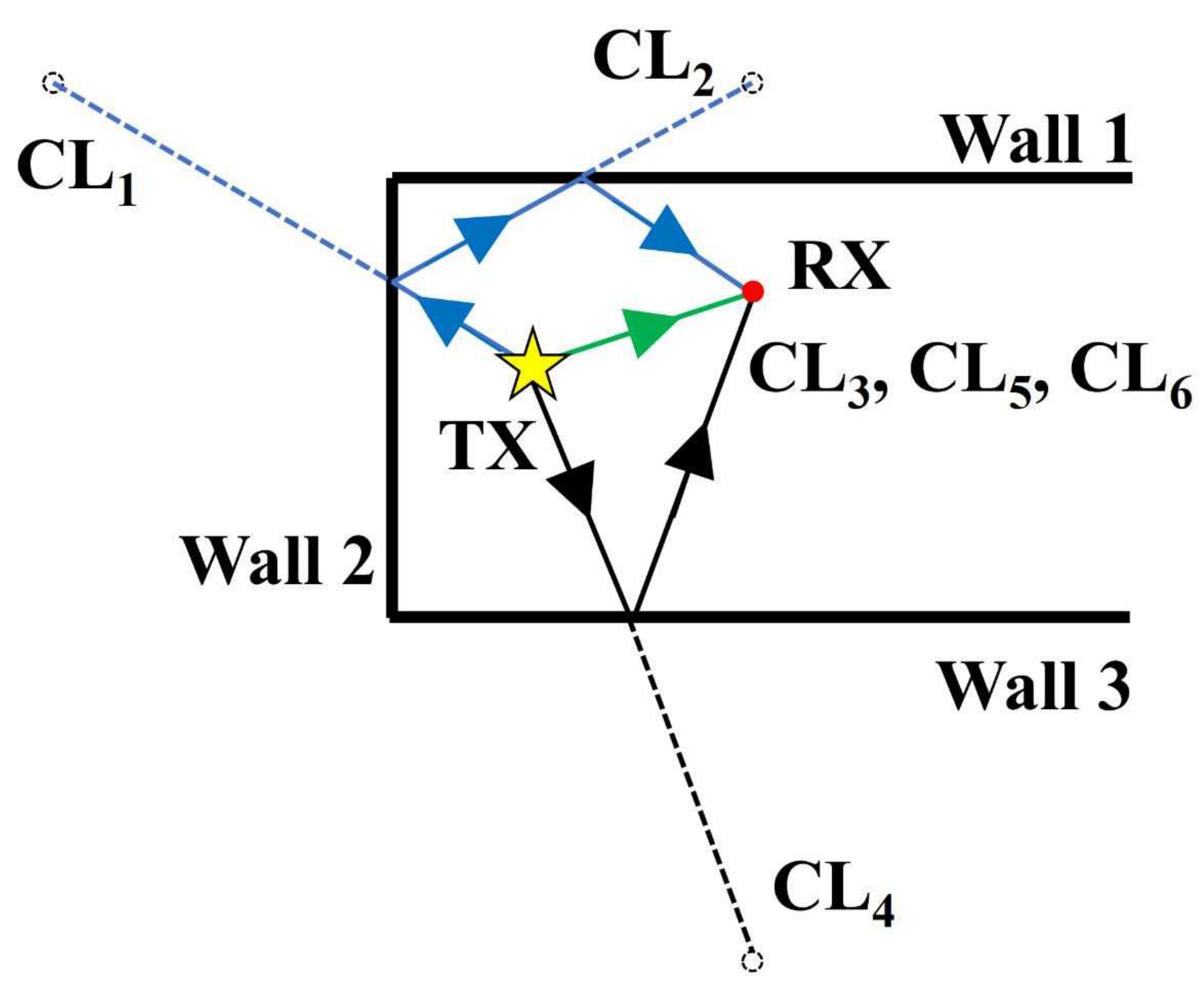}
	\caption{ Three multipath components arrive at the user (RX) shown above - one LOS component (in green) and two NLOS components (in blue and black). Of the six candidate locations for the user, based on AoD and ToF measurements at the BS ($ \text{CL}_1 - \text{ CL}_6 $), three candidate locations $ (\text{CL}_3,\text{ CL}_5, \text{ CL}_6) $ correspond to the actual location of the user. The position of the user is estimated to be the modal candidate location (i.e. $ \text{CL}_3,\text{ CL}_5, \text{ CL}_6 $). }
	\label{fig:candidates}
\end{figure}

Consider the case when there is at most one reflection or transmission of the signal before it reaches the user. If the ToF and AoD of a multipath signal sent from the BS is known, there are two possible locations of the user, as seen in Fig. \ref{fig:config_1}. If the ray reached the user after one reflection, the user and BS must lie on the same side of the reflecting object. If the ray reached the BS directly from the user, or through one obstruction, the BS and user must lie on opposite sides of the obstruction. The possible locations of the user, based on ToF and AoD at the BS shall henceforth be referred to as \textit{candidate locations}. When the signal is reflected or transmitted multiple times, each successive reflection/transmission creates more candidate locations, as seen in Fig. \ref{fig:candidates}. The process of finding candidate locations is repeated for all multipath components. If a single multipath component is received by the user, the BS will not be able to determine which candidate location corresponds to the user's true location. However, when two or more multipath components arrive at the user, a majority of the candidate locations will correspond to the true location of the user. For each multipath component arriving at the user, one candidate location calculated based on the AoD and ToF of the multipath component corresponds to the true user's location. Fig. \ref{fig:candidates} depicts all the candidate locations when three multipath components are received by the user (RX) from the BS (TX). The user's location corresponds to the candidate location identified by the maximum number of multipath components.

Errors in ToF measurements cause the BS to incorrectly estimate the path length to the user. Due to inaccurate AoD measurements the BS incorrectly estimate the user’s bearing. As a result, candidate locations estimated by MAP-AT using imprecise ToF and AoD information will not coincide with the user’s true location. However, it is likely that the candidate locations will be close to the user’s true location, and hence close to one-another. MAP-AT is modified to first group the candidate locations that are close to one another (at a distance of up to $ d =$ 40 cm, where $ d $ is a tunable parameter) to form candidate location clusters. The user position is estimated to be the centroid of the candidate location group with maximal members.

MAP-AT may be generalized to use multiple BSs in a straightforward fashion. Candidate locations are determined, corresponding to all the multipath components received by the user from all BSs. MAP-AT then proceeds in a similar fashion to the case when only one BS was utilized, by finding the candidate location identified by the maximum number of multipath components.

\subsection{Configuration-\Romannum{2} - user in the transmission mode }
In configuration-\Romannum{2}, the user transmits a known signal to the BS. The BS may calculate the AoA of each multipath signal that reaches the BS. The ToF of the each multipath component from the user arriving at the BS may either be estimated via the round trip time of the multipath component or the one-way propagation time. If the ToF is estimated via round-trip time, the BS does not need to send the ToF of the multipath components to the user via a feedback channel. Additionally, synchronization between the user and the BS is not required. The BS only needs to send the AoA of each multipath component to the user. Candidate locations of the RX are found analogously to configuration-\Romannum{1}. The user's location corresponds to the candidate location identified by the maximum number of multipath components.
\subsection{Computational Complexity}
MAP-AT first calculates the candidate locations after which close-by candidate locations are grouped together. Let there be $ M $ multipath components arriving the RX, after $ k $ reflections and/or transmissions. Assuming at most $ k=3  $ reflections and/or transmissions, MAP-AT must calculate at most $ M \cdot 2^k  = 8\cdot M$ reflections to locate all candidate locations. Let there be $ n $ distinct candidate locations. Grouping all $ N $ candidate locations will involve at most ${n\choose 2}$ euclidean distance calculations. Thus, the computational complexity of MAP-AT is $ \mathcal{O}( n^2). $
\section{Simulations and Results}\label{sec:simulations}
Simulations of localization using MAP-AT were conducted at 73 GHz by synthesizing ToF and AoD measurements at 100 TR combinations via NYURay, of which 79 were in NLOS and 21 were in LOS. BS locations TX 1-5 were identical to the BS locations chosen in the indoor propagation measurement campaign conducted at the NYU WIRELESS research center on the 9th floor of 2 MetroTech Center in downtown Brooklyn, New York \cite{Mac15b}. Additional BSs, TX 6 and 7 were added to the simulations to ensure adequate coverage throughout the floor. 100 user locations were chosen uniformly at random over the entire floor. The research center is a typical large office, with cubicles, walls made of drywalls and windows. A map of the indoor propagation environment is shown in Fig \ref{fig:Indoor_locations}. Since localization accuracy does not depend on the user configuration (\Romannum{1} or \Romannum{2}), without loss of generality, configuration-\Romannum{1} was chosen to analyze the performance of the positioning algorithm.

The position of each user was determined using a single BS. To make the simulations realistic, zero mean Gaussian noise with standard deviation $ \sigma_{AoD} = 0.5^\circ$ was added to the AoD measurements. Three difference levels of Gaussian noise were added to ToF measurements. The standard deviation $ \sigma_{ToF} $ was set to 0.25 ns, 0.5 ns and 1 ns. The positioning error for each user was defined to be equal to the 3-D Euclidean distance between the position estimate and the true position of the user. The rms positioning error for each user was calculated over 100 simulations at all three ToF measurement noise levels.
\subsection{Localization Performance with one BS}
With noise levels of $ \sigma_{AoD} $ =  0.5\textdegree\ and $ \sigma_{ToF} $ = 0.25 ns, the mean rms positioning error was 12.6 cm in LOS conditions and 16.3 cm in NLOS conditions, over a total of 100 user locations. Table \ref{table:one_BS} illustrates how the localization error varies with TR separation distance in LOS and NLOS environments. Increasing $\sigma_{ToF} $ to 1 ns degraded the performance of the algorithm. The median localization accuracy was 20.1 cm, with a mean localization error of 50.0 cm.

\begin{table}[]
	\centering
	\caption{Performance of the map-assisted localization algorithm for different TR separation distances in LOS and NLOS environments. }\label{table:one_BS}
	\begin{tabular}{|c|c|c|c|}
		\hline
		TX-RX distance & Env.                                                 & Number of Users                                  & \begin{tabular}[c]{@{}c@{}}Mean Localization \\   error (cm)\end{tabular} \\ \hline
		\textless 10 m & \begin{tabular}[c]{@{}c@{}}LOS\\   NLOS\end{tabular} & \begin{tabular}[c]{@{}c@{}}16\\   58\end{tabular} & \begin{tabular}[c]{@{}c@{}}12.6\\   16.3\end{tabular}                      \\ \hline
		10-25 m        & \begin{tabular}[c]{@{}c@{}}LOS\\   NLOS\end{tabular} & \begin{tabular}[c]{@{}c@{}}5\\  21\end{tabular}  & \begin{tabular}[c]{@{}c@{}}31.8\\  38.8\end{tabular}                      \\ \hline
		(all)      & \begin{tabular}[c]{@{}c@{}}LOS\\   NLOS\end{tabular} & \begin{tabular}[c]{@{}c@{}}21\\   79\end{tabular}  & \begin{tabular}[c]{@{}c@{}}17.2\\   22.3\end{tabular}                     \\ \hline
	\end{tabular}
	\vspace{-1.5em}
\end{table}
\subsection{Localization Performance with multiple BS}
Out of the 100 users, 81 were within the range of at least two BS. Additionally, 50 users were within the range of three BSs. The performance of the map-assisted localization algorithm with more than one BS is summarized in Table \ref{table:two_BS}. For the users not in LOS of any BS, increasing the number of BSs used for localization helps reduce the localization error. When one BS is used to localize a NLOS user, an average localization error of 22.3 cm is achieved. With two and three BSs, the localization error dropped to 15.2 cm and 5.5 cm respectively. The localization error for users in LOS remains constant ($\sim$ 15 - 20 cm).

\begin{figure}
	\centering
	\includegraphics[width=0.48\textwidth]{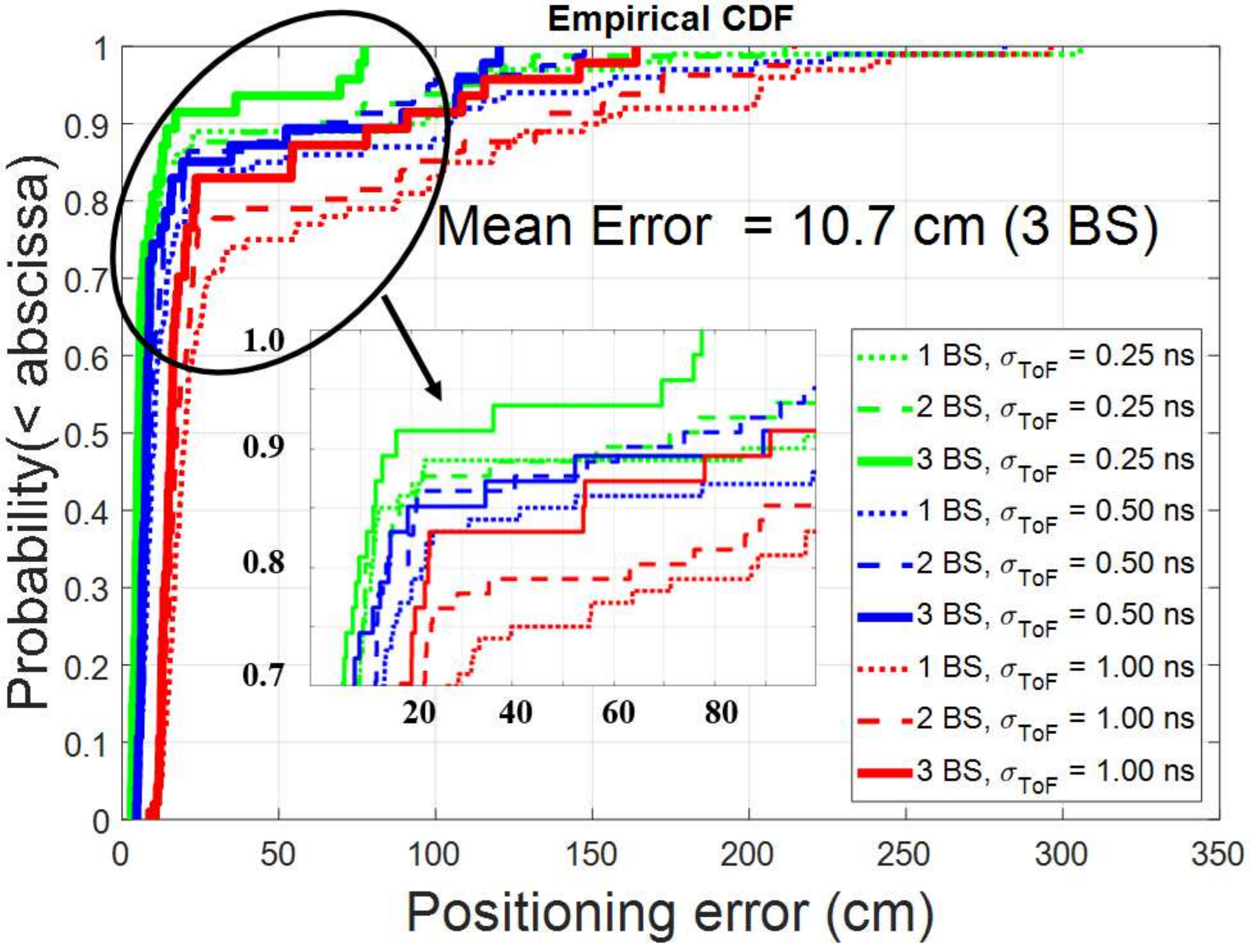}
	\caption{ The CDF of the positioning error, when a user is localized using one, two, and three BSs, with $ \sigma_{AoD} =$ 0.5\textdegree. }
	\label{fig:positioning_error_candidates}
\end{figure}

The cumulative density functions (CDFs) of the positioning errors, with one, two, and three BSs are provided in Fig. \ref{fig:positioning_error_candidates}. With a single BS, the 90 percentile rms error was 97 cm. A maximum rms error of 3.05 m was observed, when the user was located behind a corner, due to which a small error in AoD estimation led to incorrect prediction of the reflecting obstruction. When three BSs were utilized, all users were localized to within 80 cm.

\begin{table}[]
	\centering
	\caption{Performance of the map-assisted localization algorithm with more than one BS. }\label{table:two_BS}
	\begin{tabular}{|c|c|c|}
		\hline
BS-User	Link Type                                                 & Number of Users                                  & \begin{tabular}[c]{@{}c@{}}Mean Localization \\   Error (cm)\end{tabular} \\ \hline
		1 LOS, 1 NLOS  & 38 & 22.9                     \\ \hline
		2 NLOS & 43  & 15.2                    \\ \hline
		1 LOS, 2 NLOS  & 29 & 14.5                     \\ \hline
		3 NLOS & 21  & 5.5                      \\ \hline
		
	\end{tabular}
	\vspace{-1.5em}
\end{table}

\section{Conclusion}\label{sec:conclusion}
This paper has introduced MAP-AT, a novel method for map-assisted data fusion of AoD and ToF information, to provide centimeter-level localization. A 3-D map of the environment (generated on-the-fly by the user or downloaded a-priori), in concert with AoD and ToF information may be used to predict the propagation path of multipath components received by the user. MAP-AT was tested using NYURay, a 3-D ray tracer developed herein. The ray tracer was calibrated based on real-world measurements conducted at 28 and 73 GHz\cite{Mac15b} and predicted the received power accurately to within 6.5 dB. Based on simulations conducted at 100 uniformly distributed user locations, a mean localization accuracy of 17.2 cm in LOS and 22.3 cm in NLOS was achieved in a typical large office environment using a single BS per user, with TR separation distances varying from 1.5 m to 24.5 m. By using three BSs, the localization error for LOS users remained the same while the localization error for NLOS users dropped to 5.5 cm. Future work will study the effects of using inaccurate building maps on the position locationing accuracy. Map-assisted outdoor localization accuracy, via a data fusion of AoA, ToF, and power, will also be investigated.

\section{Acknowledgments}
This material is based upon work supported by the NYU WIRELESS Industrial Affiliates Program and National Science Foundation (NSF) Grants: 1702967 and 1731290.
\bibliographystyle{IEEEtran}
\bibliography{GC_19_references}{}

\end{document}